\documentclass[12pt, amsmath, color]{article}
\usepackage{mathrsfs,extarrows}
\usepackage{url}
\usepackage{float}

\usepackage{longtable}
\usepackage{caption}
\usepackage[colorlinks=true,citecolor=blue,linkcolor=blue]{hyperref}%

\usepackage{authblk}

\makeatletter
\def\singlespace{\def\baselinestretch{1}\@normalsize}

\newcommand{\bea}{\begin{eqnarray}}
\newcommand{\eea}{\end{eqnarray}}
\newcommand{\bbe}{\begin{eqnarray*}}
\newcommand{\ebe}{\end{eqnarray*}}
\newcommand{\bn}{\begin{enumerate}}
\newcommand{\en}{\end{enumerate}}
\newcommand{\bcor}{\begin{corollary}}
\newcommand{\ecor}{\end{corollary}}

\newcommand{\blem}{\begin{lemma}}
\newcommand{\elem}{\end{lemma}}

\newcommand{\bprop}{\begin{proposition}}
\newcommand{\eprop}{\end{proposition}}

\newcommand{\beq}{\begin{equation}}
\newcommand{\eeq}{\end{equation}}
\newcommand{\beqn}{\begin{eqnarray}}
\newcommand{\eeqn}{\end{eqnarray}}
\newcommand{\beqnn}{\begin{eqnarray*}}
\newcommand{\eeqnn}{\end{eqnarray*}}

\newtheorem{thm}{Theorem}
\newtheorem{lem}{Lemma}
\newtheorem{rem}{Remark}
\newtheorem{cor}{Corollary}

\newcommand{\bel}[1]{\begin{equation}\label{#1}}
\newcommand{\Thm}[1]{\begin{thm}\label{thm.#1}}
\newcommand{\Cor}[1]{\begin{cor}\label{cor.#1}}
\newcommand{\Prop}[1]{\begin{proposition}\label{prop.#1}}
\newcommand{\Lm}[1]{\begin{lem}\label{lm.#1}}
\newcommand{\Ass}[1]{\begin{assumption}\label{ass.#1}}
\newcommand{\Ex}[1]{\begin{example}\label{ex.#1}\rm}
\newcommand{\Def}[1]{\begin{definition}\label{def.#1}}
\newcommand{\Rem}[1]{\begin{rem}\label{rem.#1}}

        \def\b1{{\bf 1\!\!\!1}}  % Letter + number

\usepackage{JASA_manu}
\usepackage{array}
\usepackage[dvips]{graphicx}
\usepackage{amsmath,color}
\usepackage{amssymb,enumerate}
\usepackage{natbib,epsfig,graphicx,rotating}
\usepackage[ruled,vlined]{algorithm2e}

\numberwithin{equation}{section}
\def\boxit#1{\vbox{\hrule\hbox{\vrule\kern6pt
          \vbox{\kern6pt#1\kern6pt}\kern6pt\vrule}\hrule}}

\newcommand{\bbi}{\begin{itemize}
 \itemsep -.1in}
\newcommand{\eei}{\end{itemize}}

\title{\bf The Effects of Stringent Interventions for Coronavirus Pandemic}

\begin{document}

\def\spacingset#1{\renewcommand{\baselinestretch}%
{#1}\small\normalsize} \spacingset{1}

\title{\bf The Effects of Stringent Interventions for Coronavirus Pandemic} %\protect\thanksref{T1}}
\author[a,1]{Ting Tian}
\author[a,1]{Wenxiang Luo}
\author[a,1]{Yukang Jiang}
\author[a]{Minqiong Chen}
\author[b]{Canhong Wen}
\author[a]{Wenliang Pan}
\author[b,2]{Xueqin Wang}
\affil[a]{School of Mathematics, Sun Yat-sen University}
\affil[b]{School of Management,University of Science and Technology of China}

\date{}                     %% if you don't need date to appear
\setcounter{Maxaffil}{0}
\renewcommand\Affilfont{\itshape\small}
 \maketitle

\renewcommand{\baselinestretch}{1.5}

\begin{footnotetext}[1]
{Ting Tian, Wenxiang Luo and Yukang Jiang contributed equally to this article }
\end{footnotetext}
\begin{footnotetext}[2]
{Xueqin Wang: wangxq20@ustc.edu.cn.}
\end{footnotetext}
%%%%%%%%%%%%%%%%%%%%%%%%%%%%%%%%%%%%%%%%%%%%%%%%%%%%%%%%%%%%%%%%%%%%%%%%%%%%%

\date{}
\maketitle

\begin{abstract}
\baselineskip=20pt

The pandemic of COVID-19 has caused severe public health consequences around the world. Many interventions of COVID-19 have been implemented. It is of great public health and societal importance to evaluate the effects of interventions in the pandemic of COVID-19. In this paper, with help of synthetic control method, regression discontinuity and a Susceptible-Infected and infectious without isolation-Hospitalized in isolation-Removed (SIHR) model, we evaluate the horizontal and longitudinal effects of stringent interventions implemented in Wenzhou, a representative urban city of China, where stringent interventions were enforced to curb its own epidemic situation with rapidly increasing newly confirmed cases. We found that there were statistically significant treatment effects of those stringent interventions which reduced the cumulative confirmed cases of COVID-19. Those reduction effects would increase over time. Also, if the stringent interventions were delayed by 2 days or mild interventions were implemented instead, the expected number of cumulative confirmed cases would have been nearly 2 times or 5 times of the actual number. The effects of stringent interventions are significant in mitigating the epidemic situation of COVID-19. The slower the interventions were implemented, the more severe the epidemic would have been, and the stronger the interventions would have been required.

\end{abstract}

\bigskip

 \baselineskip=22pt

\newpage

\section{Introduction}

The novel coronavirus disease (COVID-19) as an acute infection has rapidly spread over 200 countries in the world since December 2019 \citep{world2020rolling}. There were 2,804,799 reported cumulative confirmed cases and 193,711 cumulative deaths around the world and 84,341 cumulative confirmed cases and 4,643 cumulative deaths from China until April 26 , 2020 \citep{CDC2020}. These numbers are rapidly arising daily. This coronavirus family includes the virus that caused to Severe Acute Respiratory Syndrome (SARS) in 2003 and another that caused to Middle East Respiratory Syndrome (MERS) in 2012 \citep{zhu2020novel}. Early cases of COVID-19 suggested that it might not be as severe as SARS-CoV and MERS-CoV, but the rapid increase in confirmed cases and the evidence of human-to-human transmission indicated that this coronavirus was more contagious than both SARS-CoV and MERS-CoV \citep{wang2020novel,guan2020clinical,chan2020familial}.

Compared with the outbreak of SARS, the Chinese government mounted responses rapidly to contain COVID-19 by reducing the risk of personal exposure and transmission \citep{tu2020epidemic}. As the city with the largest number of confirmed cases in Zhejiang province \citep{CDC2020CN}, where it activated the first-level public health emergency response on January 23, 2020, Wenzhou had suspended all inter-provincial and inter-city shuttle buses, and chartered buses, and carried out temperature testing for passengers of buses arriving in Wenzhou as its own policies to deal with the epidemic situation since January 27, 2020 \citep{Wenzhou03}. These strict control measures were implemented across Wenzhou, where one family member was assigned to go out for necessity purchase of each household every two days since February 1, 2020 \citep{Wenzhou06}. Thus, Wenzhou had adopted the stringent intervention policies, including both traffic blockade inside and outside the city, and the strictly closed management of the community, to intensively prevent and control of the epidemic situation of COVID-19. In addition, an industrial city of Guangdong province, which activated the first-level public health emergency response on January 23, 2020 \citep{Guangdong}, Shenzhen adopted closely following the close contact of confirmed cases and isolating them on the same time. Comparatively, Shenzhen implemented relatively mild but early interventions.

This paper concentrated on the strict intervention policies of COVID-19 that started on February 1, 2020 in Wenzhou and evaluated the treatment effects of those strict interventions, providing insights into the effects of the strict interventions. We also assessed what could have happened in Wenzhou if the stringent interventions were delayed by 0 to 5 days or if mild interventions were implemented instead. We accomplished this through simulation by using the time-varying Susceptible-Infected and infectious without isolation-Hospitalized in isolation-Removed (SIHR) model \citep{tan2020}. Finally, we simulated the likely outcomes if relatively mild interventions as used in Shenzhen, China, were implemented in lieu of the stringent interventions.

In the next section, we describe the classical causal effects methods and dynamic transmission models of COVID-19 in detail. The corresponding analysis of epidemic data in Wenzhou are illustrated in Section 3. Section 4 discusses the effects of stringent interventions of COVID-19.

\section{Methodologies}
\subsection{Data collection}

We compiled a total of 504 confirmed cases \citep{Health} from January 21, 2020 when the confirmed cases were first reported, to February 25, 2020 when no newly reported confirmed cases had been reported for a week, from the official website of the health commission of Wenzhou, and obtained the residential population of Wenzhou from the 2019 statistical yearbook of the Zhejiang Provincial Bureau of Statistics \citep{SBZhejiang}. The total confirmed cases of Shenzhen during the same period were collected. The cases per 100,000 individuals were calculated by the prevalence of COVID-19 multiplying 100,000 individuals in Wenzhou. We searched the daily number of cumulative confirmed cases in each city of China using the R package "nCov2019" \citep{wu2020open}, and collected the number of cumulative confirmed cases from the top 100 cities for the epidemic of COVID-19 (except Hubei province) on February 25, 2020. The corresponding population density and GDP per capita statistics of these 100 cities were obtained from their latest statistical yearbooks.

\subsection{Synthetic Control Method}

To identify the treatment effects of the highly stringent interventions implemented on February 1, 2020, we compared Wenzhou with cities without those interventions. The synthetic control method was used to construct a data-driven control-group in the absence of those interventions in Wenzhou \citep{abadie2003economic,abadie2010synthetic}. The observed cases per 100,000 for Wenzhou (active intervention city) at the time $t$ with interventions were denoted by $Y_{1t}^1$, and cases per 100,000 for cities $j$ at the time $t$ without interventions by $Y_{jt}^0$. The difference of $Y_{1t}^1-Y_{1t}^0$ was the treatment effect of the interventions on the cases per 100,000 for the active intervention city in the post-interventions period. Note that $Y_{1t}^0$ is counterfactual for the active intervention city in the post-interventions period. Reliable estimates of cases per 100,000 for a synthetic intervention city were used to derive $Y_{1t}^1-Y_{1t}^0$, where a synthetic intervention city resembled the active intervention city by possible pre-interventions characteristics. We selected the GDP per-capita \citep{zhang2020epidemic}, population density \citep{oto2020regional}, and three-day cases per 100,000 before the implementation of interventions to construct a resembling city of Wenzhou. The effects of unobserved factors varied over time and were controlled by the linear combination of cases per 100,000 of pre-interventions \citep{abadie2010synthetic}. Thus, the treatment effect of interventions in Wenzhou was formulated as \citep{abadie2011synth}:
\begin{align}
Y_{1t}^1-Y_{1t}^0=Y_{1t}^1-\sum_2^{J+1}w_j^* Y_{jt}^0. \label{e2}
\end{align}

To determine $J$, the number of cities demographically similar to Wenzhou, we selected the top 100 cities of cumulative confirmed cases on February 25, 2020 in China (except Hubei province) and used hierarchical clustering \citep{ward1963hierarchical} to group those cities based on population density and GDP per capita. A homogenous group including Wenzhou and other $J$ cities were grouped together. Thus, those $J$ cities could be considered as ¡°highly resembling¡± to Wenzhou. The sum of weights for each of $J$ cities was 1 and the values of weights in Equation (\ref{e2}) were determined by pre-interventions characteristics. The placebo tests were used to identify the magnitude of the treatment effect (i.e. the causal effect of interventions) \citep{abadie2010synthetic}.

\subsection{Regression Discontinuity}

For the strict intervention policies of Wenzhou, we vertically compared the difference in the epidemic situation of COVID-19 before and after the implementation of such intervention policies. Regression Discontinuity (RD) \citep{thistlethwaite1960regression,hahn2001identification,imbens2008regression} was used to evaluate whether the implementation of strict intervention policies of Wenzhou on February 1, 2020 had a significant effect on its own COVID-19 epidemic. The outcome of these stringent interventions was
formulated as:
\begin{align}
\log y_{t} =\alpha_0+\alpha_1 h_t+\alpha_2 (t-c)+\alpha_3 h_t (t-c)^k+\varepsilon_t \label{e1}
\end{align}
where the logarithmic transformation of the daily number of cumulative confirmed cases as a response variable. The values of $t$ include 1 to 36, representing the number of days starting from January 21, 2020 as a rate variable in Equation (\ref{e1}) of RD, for example, the date of February 1 was denoted as 12 of $t$. $h$ is an indicator variable for the implementation of the strict intervention policies in Wenzhou, given a value of 1 as the implementation of the policies but a value of 0 as no implementation of the policies. This indicates that the values are 0 and 1 before and after February 1, 2020 (including February 1 itself), respectively and the subscript of $t$ indicates the number of days as defined before. $c$ is the number of days corresponding to the cutoff-point whose value is 12, representing the policy implemented on February 1, 2020. The values of $k$ could be denoted as the $k^{th}$ order of interaction effect between rate variable and policy. We compared the $k^{th}$ order of interaction effect item in the Equation (\ref{e1}) by AIC.

In RD, to determine whether $c$ is an obvious jump point, the plot of both the rate variable and response variable was used to identify their relationship \citep{calonico2015optimal}. Different numbers of bins could be used to divide both the left-hand side and right-hand side of the cutoff-point to small intervals. Thus, the scatter plot of both median values of the rate variable and mean values of the response variable could be drawn \citep{calonico2015rdrobust}. The actual rate variables are weighted uniformly in each bin, and the fitted curves of both the left-hand side and right-hand side are used to examine whether the cutoff-point can be considered as a threshold.

\subsection{Susceptible-Infected and infectious without isolation-Hospitalized in isolation-Removed (SIHR) model}

The dynamic system of SIHR with four classes: Susceptible ($S$), infected and infectious without isolation ($I$), hospitalized in isolation ($H$), removed ($R$) \citep{li2018introduction, hsieh2004sars,tan2020} was defined as:
\begin{align}
& \frac{\mathrm{d} S(t) }{\mathrm{d} t}  =- \alpha \frac{I(t)}{N} S(t),\nonumber\\
& \frac{\mathrm{d} I(t) }{\mathrm{d} t} =\alpha {\frac{I(t)}{N} S(t)}-\beta I(t),\nonumber\\
& \frac{\mathrm{d} H(t)}{\mathrm{d} t}  = \beta I(t)-\gamma H(t) , \\
& \frac{\mathrm{d} R(t) }{\mathrm{d} t} = \gamma H(t) ,\nonumber\\
& N=S(t)+I(t)+H(t)+R(t), \nonumber
\end{align}
where the transmission rate $\alpha$ is the average rate of being infected given contact over unit time, $1/\beta$ is the mean of the incubation period, $1/\gamma$ is the mean of the hospitalization period. To consider the effects of interventions, we introduced the time-varying transmission rate, which was defined as \citep{tan2020}:
\begin{align}
\alpha_{\alpha_0,d,m}(t)=\frac{\alpha_0}{1+\exp (\lambda_m(t-d-m/2))},
\end{align}
where $\alpha_0$ denotes the maximum transmission rate of COVID-19 during the early outbreak, $d$ represents the time when the interventions start to be effective and the transmission rate starts to decline, $m$ is the duration of a process where the epidemic is to nearly vanish, $\lambda_m$ is selected as $(2\log ((1-\varepsilon)/\varepsilon))/m$ and $\varepsilon$ is specified to be 0.01. The smaller the values of $d$ and $m$, the earlier effectiveness and the stronger intensity of interventions were implemented, respectively. We simulated the likely outcomes of delaying stringent interventions from 0 to 5 days by changing the values of $d$ and of mild interventions on the same starting time of Wenzhou policies by changing the value of $m$ to Shenzhen policies.

\section{Case Study}

\subsection{Synthetic Control Method Analysis}

We selected the top 100 cities outside Hubei province according to the cumulative confirmed cases on February 25, 2020 (\textbf{Table \ref{table3}} in the Supplementary Materials). Based on the per capita GDP and population density of each city, these cities (including Wenzhou) were clustered into 4 groups using hierarchical clustering. Among them, a homogenous group consisted of Wenzhou and other 45 cities (\textbf{Figure \ref{fig5}} in the Supplementary Materials), where Taizhou of Zhejiang province implemented similar policies to Wenzhou and was excluded from our analysis. The remaining 44 cities formed a ¡°counterfactual¡± city resembling Wenzhou (\textbf{Table \ref{table4}} in the Supplementary Materials).

\begin{figure}[H]
\begin{center}
\includegraphics[totalheight=5.4in, width=6.8in, origin=c]{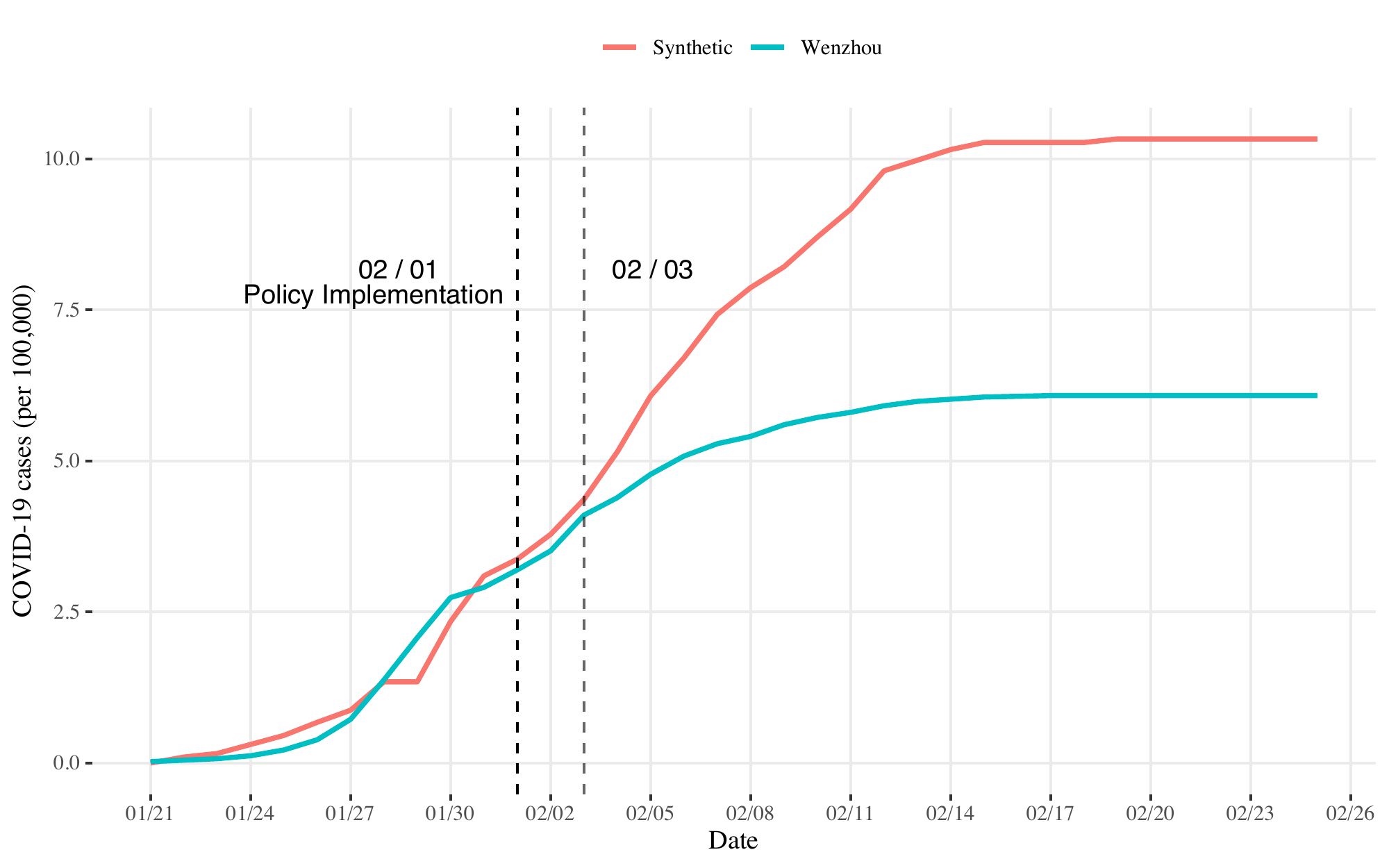}
\end{center}
\caption{\textbf{The trends of COVID-19 cases per 100,000 between Wenzhou and Synthetic Wenzhou from January 21 to February 25, 2020.} The black dashed line indicated the time of implementation of the intervention policies in Wenzhou. The grey dashed line indicated 2 days delay of implementation of the intervention policies in Wenzhou.} \label{fig2}
\end{figure}

Before the strict interventions were implemented on February 1, 2020, the trends of actual cases per 100,000 in both Wenzhou and "synthetic Wenzhou" were highly similar, suggesting that this synthetic city can be used to estimate the "counterfactual" results of Wenzhou. After 2 days of the implementation of the policies, the growth rate of actual cases per 100,000 in Wenzhou remarkably slower than that of "synthetic Wenzhou". The number of cases per 100,000 in "Synthetic Wenzhou" on February 25, 2020 was 10.32, which is 1.69 times the actual Wenzhou (6.08) (\textbf{Figure \ref{fig2}}). In other words, Wenzhou had not implemented the strict interventions on February 1, 2020, the number of cumulative confirmed cases would have reached 954 on February 25, 2020, i.e., the epidemic of COVID-19 in Wenzhou would have expanded to approximately 1.7 times.

\begin{figure}[H]
\begin{center}
\includegraphics[totalheight=5.4in, width=6.8in, origin=c]{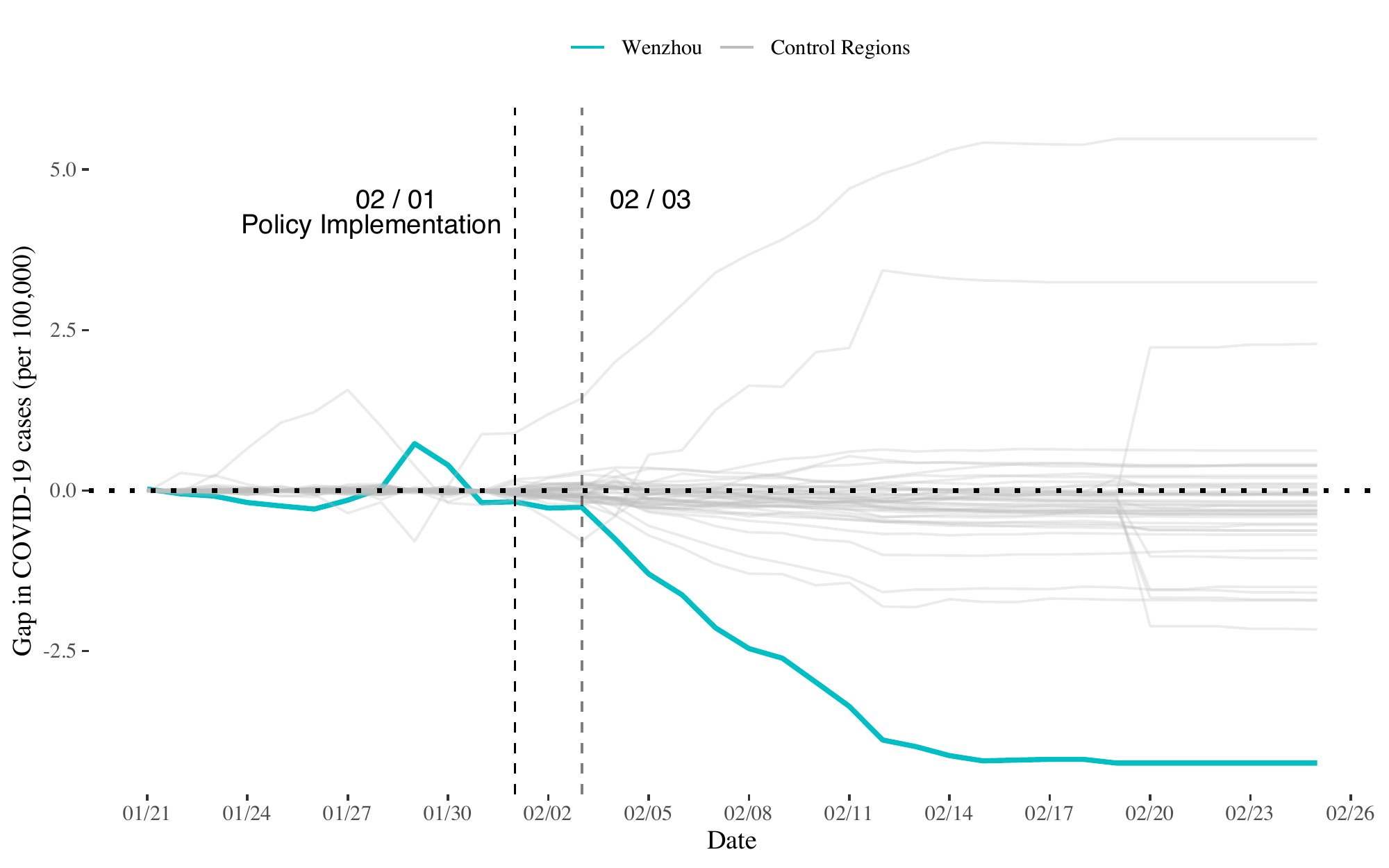}
\end{center}
\caption{\textbf{The permutation test of the treatment effect of the implementation of the policies in Wenzhou and other 44 control regions.} All grey curves represent the placebo tests of COVID-19 cases per 100,000 gaps between a control city (an arbitrary random city of 44 cities) and synthetic control city (a combination of remaining 43 cities and Wenzhou), and the blue curve represents the placebo test of COVID-19 cases per 100,000 gap between Wenzhou and synthetic Wenzhou.} \label{fig3}
\end{figure}

A placebo test was performed to determine the significance level of the difference in the trends of COVID-19 cases per 100,000. To this end, we plotted the gap curves between Wenzhou and synthetic Wenzhou by in turn exchanging Wenzhou and one of each 44 cities in the homogenous group of Wenzhou. The gap of COVID-19 cases per 100,000 between Wenzhou and our synthetic Wenzhou was the largest compared to the rest, i.e. the negative effect of the intervention policies on COVID-19 per 100,000 in Wenzhou was the lowest of all. For those 44 cities, the probability of having a gap for Wenzhou under a random permutation of the control measures was $5\%$, which is conventionally regarded as statistically significant. This suggested that the effect of the implementation of the policies in Wenzhou was significantly different from the implementation of the policies in the remaining 44 cities, indicating that the strict interventions of Wenzhou might have significantly reduced the COVID-19 cases per 100,000 (\textbf{Figure \ref{fig3}}).

\subsection{Regression Discontinuity Analysis}
The values of AIC for each regression discontinuity model for different $k^{th}$ order of interaction effects between rate variable and policy were shown in \textbf{Table \ref{table0}}.
\begin{table}[H]
\caption{The values of AIC for each model with different $k^{th}$ order of interaction effects}
\vspace*{1\baselineskip}
\centering
\begin{tabular}{ccc}
 \hline
  Models & $df$ & AIC \\
  \hline
  No interaction & 4 & 97.344 \\
  First order & 5 & -28.861  \\
  $Time^2$ without interaction & 5 & 44.057\\
  Second order & 6 & -40.846 \\
  $Time^3$ without interaction& 6 & -34.639 \\
  Third order & 7 & -42.019 \\
 \hline
\end{tabular}\label{table0}
\end{table}

\begin{figure}[H]
\begin{center}
\includegraphics[totalheight=5.4in, width=5.6in, origin=c]{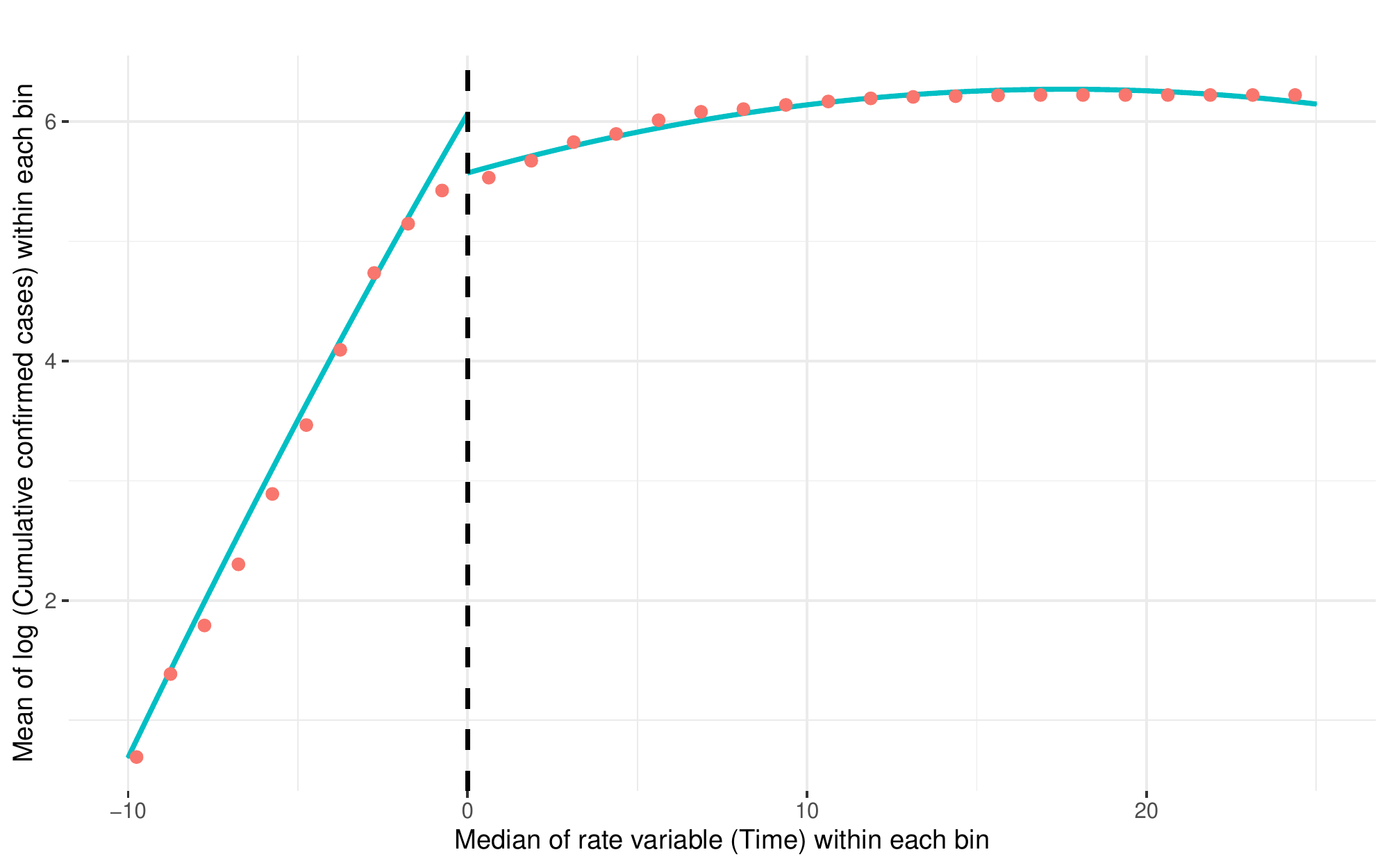}
\end{center}
\caption{\textbf{The plot of both rate variable and response variable}. The green fitted curves of relationship
between median of time within each bin and mean of log(cumulative confirmed cases) within
each bin, the black dashed vertical line represents the time of cutoff-point (with value 12)} \label{fig1}
\end{figure}

Looking at \textbf{Table \ref{table0}}, the model included item $Policy$*$Time^3$ ($h_t(t-c)^3$) had the lowest value of AIC but it had more parameters ($df=7$). Also, the slopes of the fitted curves between the left-hand side and the right-hand side of the cutoff-point were different (\textbf{Figure \ref{fig1}}). The points lied on the right-hand side of quadratic curve. Thus, RD could be used to examine the treatment effect of interventions implemented in Wenzhou and the model included the second order of interaction effects between policy and rate variable was reasonable. The coefficient of interventions is -0.350 (p-value: 0.003) and the interaction effect between interventions and time is -0.438 (p-value: $<$0.001) indicating that there was a significant treatment
effect of highly stringent interventions implemented in Wenzhou on February 1, 2020 (\textbf{Table \ref{Table1}}).

\begin{table}[H]
\caption{The coefficients of variables in the RD and their corresponding p-values}
\centering
\vspace*{1\baselineskip}
\begin{tabular}{cccc}
 \hline
  Variables & Coefficients & p-values \\
  \hline
  Intercept & 5.958 & $<$0.001 \\
  $Policy$ ($h_t$) & -0.350 & 0.003  \\
  $Time$ (($t-c$)) & 0.511 & $<$0.001 \\
  $Policy$*$Time$ ($h_t(t-c)$) & -0.438 & $<$0.001 \\
  $Policy$*$Time^2$ ($h_t(t-c)^2$) & -0.002 & 0.001 \\
 Adjusted $R^2$ & 0.994 &  \\
 \hline
\end{tabular}\label{Table1}
\end{table}

\subsection{Simulation of stringent interventions delay or mild interventions instead}

\begin{figure}[H]
\begin{center}
\includegraphics[totalheight=5.4in, width=6.8in, origin=c]{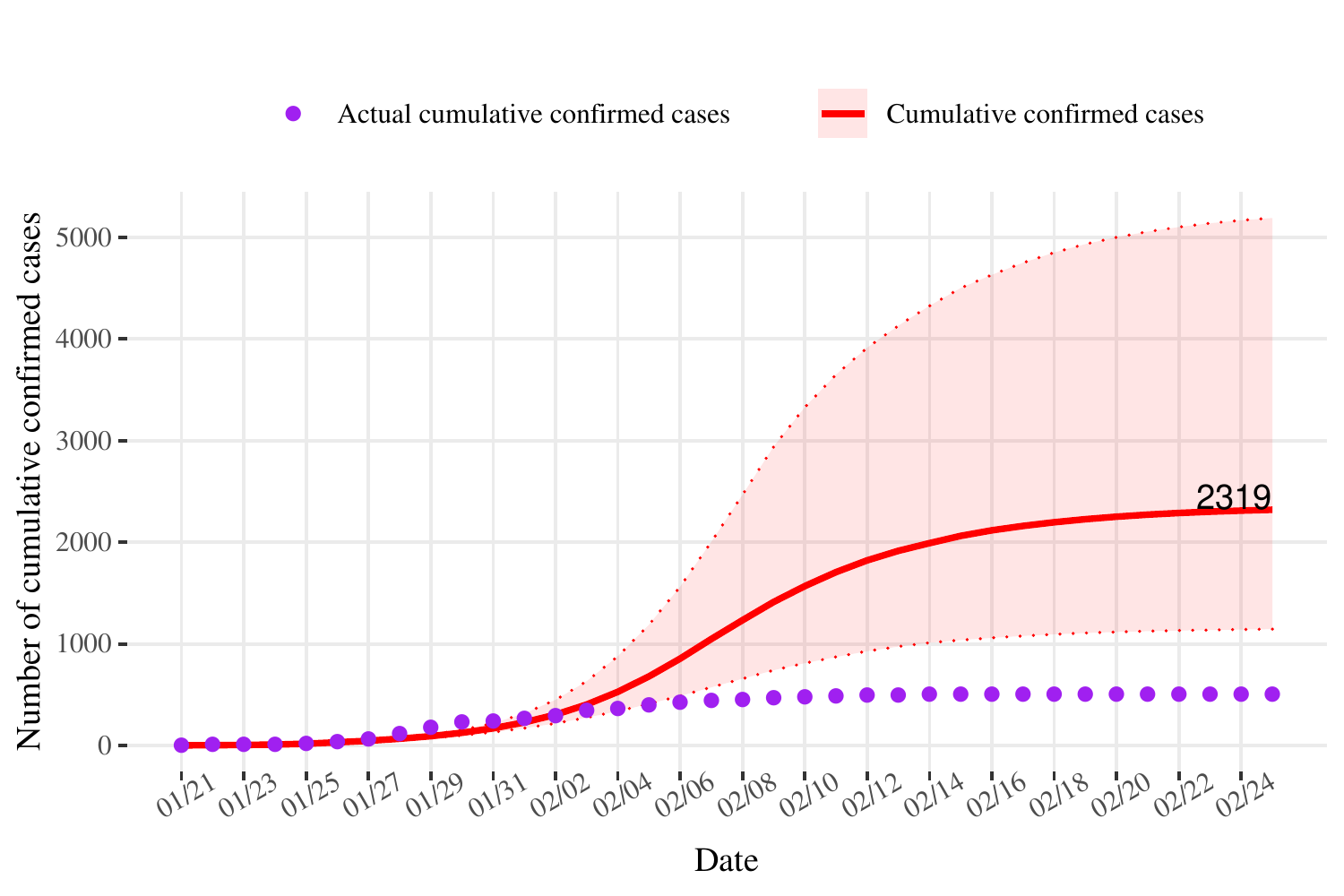}
\end{center}
\caption{\textbf{The expected number of cumulative confirmed cases for mild interventions implemented on February 1, 2020 in Wenzhou.} The points represent the actual cumulative confirmed cases of Wenzhou.} \label{fig4}
\end{figure}

Our simulation projected that the expected cumulative confirmed cases would be 1.84 times of the actual cases for a 2-day delay, 2.45 times for a 3-day delay, 3.26 times for a 4-day delay and 4.30 times for a 5-day delay on February 25, 2020. The corresponding 95$\%$ credible interval (CI) for these projected numbers of cases were presented in \textbf{Table \ref{Table2}}. The full simulation results from January 21 to February 25, 2020 are presented in \textbf{Figure \ref{fig6}} of the Supplementary Materials. According to \textbf{Table \ref{Table2}}, the expected cumulative confirmed cases for 0-day delay (i.e. the predicted cumulative confirmed cases) was very close to the actual cumulative confirmed cases of Wenzhou. The expected cumulative confirmed cases were 925 with corresponding 95$\%$ CI (571,1547) for the 2-day delay, 1233 with corresponding 95$\%$ CI (702,2233) for the 3-day delay, 1644 with corresponding 95$\%$ CI (851,3171) for the 4-day delay, and 2167 with corresponding 95$\%$ CI (1034, 4578) for the 5-day delay. The observed cumulative confirmed cases had been stable since February 17, 2020, however, the expected cumulative confirmed continued to increase if the stringent interventions were delayed. Based on \textbf{Figure \ref{fig2}}, if the stringent interventions were delayed by 2 days or more, the epidemic of COVID-19 in Wenzhou could have been more severe than that in synthetic Wenzhou (954 cumulative confirmed cases on February 25, 2020). If the mild interventions as those implemented in Shenzhen were implemented in Wenzhou on February 1, 2020, the expected number of cumulative confirmed cases would have been 2319 with corresponding 95$\%$ CI (1145,5189) on February 25, 2020 (\textbf{Figure \ref{fig4}}).

\begin{sidewaystable}[htbp]
\caption{The likely outcome of intervention implemented delay in different days including the expected cumulative confirmed cases with corresponding 95$\%$ CI from February 16 to February 25, 2020.}
\vspace*{1\baselineskip}
\centering
\begin{tabular}{cccccccc}
 \hline
  Date & Actual & 0 day & 1 day & 2 days & 3 days & 4 days & 5 days \\
  \hline
  02/16 & 503 & 514(361,749)&678(453,1059)&	892(553,1490)&1178(674,2128)&1556 (810,2998)&2034 (978,4264) \\
 02/17	&504	&517 (363,754)&	683 (456,1066)&	899 (557,1502)	&1190 (680,2151)	&1576 (819,3036)&	2064 (990,4330) \\
  02/18&	504&	519 (364,757)	&686 (458,1072)&	905 (560,1513)&	1200 (685,2170)&	1591 (826,3067)&	2087 (1000,4387) \\
 02/19	&504&521 (365,760)&	689 (460,1077)&	910 (563,1521)	&1209 (690,2185)&	1604 (832,3092)	&2106 (1008,4435)\\
 02/20	&504&	522 (366,763)&	692 (462,1081)&	914 (565,1528)	&1215 (693,2198)&	1615 (837,3112)&	2122 (1015,4472) \\
 02/21&	504	&524 (367,764)&	694 (463,1084)	&917 (567,1533)	&1220 (696,2208)	&1624 (841,3129)&	2134 (1021,4502)  \\
 02/22&	504	&525 (368,766)	&695 (464,1087)&	919 (568,1538)&	1224 (698,2217)	&1630 (845,3143)&	2145 (1025,4527)\\
 02/23&504	&525 (368,767)&	697(465,1089)&	922 (570,1542)	&1228 (700,2224)&	1636 (847,3154)	&2153 (1029,4548)\\
 02/24&	504	&526 (369,769)&	698 (465,1091)	&923 (570,1545)&	1230 (701,2229)&	1640 (850,3164)&	2161 (1032,4564)\\
 02/25&	504&526 (369,769)&	698 (466,1092)&	925 (571,1547)&	1233 (702,2233)&	1644 (851,3171)	&2167 (1034,4578)\\
  \hline
\end{tabular}\label{Table2}
\end{sidewaystable}

\section{Discussion}

We used daily reported cumulative confirmed case data from January 21 to February 25, 2020 in Wenzhou. By using the synthetic control method, the trend of COVID-19 cases per 100,000 for the ¡°synthetic Wenzhou¡± as a control group of no intervention were compared with those for Wenzhou as a treatment group with interventions. The COVID-19 cases per 100,000 in Wenzhou were significantly lower than those in ¡°synthetic Wenzhou¡± after February 3, 2020. This indicated that the implementation of the strict interventions on February 1, 2020 in Wenzhou had a significant effect on controlling the epidemic of COVID-19.

By using regression discontinuity analysis, we also concluded that the implementation of strict intervention policies had a significant treatment effect on the epidemic in Wenzhou. Moreover, the statistical treatment effects were evaluated.  That is, since the intervention policies were implemented in Wenzhou (corresponding coefficient of policy : -0.350), the growth rate of the reported cumulative confirmed cases of COVID-19 were reduced, and this "reduction" effect would increase over time (corresponding coefficient of interaction between time and policy: -0.438). Conversely, if the policies were not implemented, the reported cumulative confirmed cases of COVID-19 would increase over time (corresponding coefficient of time: 0.511).

The horizontal and longitudinal comparisons were made to examine the treatment effects of the strict intervention policies implemented in Wenzhou, including the suspension of public transportation in the city, the closure of highway junctions, and strict community access control. Since January 21, 2020, compared with other cities outside Hubei province, the number of reported cumulative confirmed cases in Wenzhou had been consistently in the top two places, i.e., its COVID-19 epidemic situation was relatively severe. If strict intervention policies were not implemented, the
outbreak of COVID-19 would be expected to expand to 1.7 times. Therefore, it can be concluded from the results of the two methods that the strict intervention policies in Wenzhou, where the epidemic was severe, had significantly suppressed its epidemic situation.

Based on our simulation, if the stringent interventions were delayed by a few days, such as 2 days, the epidemic situation would have been remarkably more severe, where the expected number of cumulative confirmed cases would have been nearly 2 times of the actual number of cases on February 25, 2020. If the mild interventions were implemented in lieu of the stringent interventions, the expected number of cumulative confirmed cases would have been 4.60 times of the actual number of cases on February 25, 2020. Overall, if the stringent interventions were delayed or the mild interventions were implemented instead, though on the same day, the expected cumulative confirmed cases would have continued to increase while the actual epidemic situation was under control in the Wenzhou.

\clearpage
\newpage

\section*{Supplementary Materials}

\setcounter{table}{0}
\begin{longtable}{p{0.18\textwidth}p{0.18\textwidth}p{0.2\textwidth}p{0.18\textwidth}p{0.18\textwidth}}
\caption{Top 100 cities of cumulative confirmed cases until 25 February (except Hubei province), and their corresponding population density and GDP per-capita}\\
\hline
\textbf{Province} &
\textbf{City}&
\textbf{Cumulative confirmed cases}&
\textbf{Population density}&
\textbf{GDP per-capita (RMB)}\\ \hline
\endfirsthead
\multicolumn{5}{c}%
{{\bfseries \tablename\ \thetable{} -- continued from previous page}} \\
\hline \textbf{Province} &
\textbf{City}&
\textbf{Cumulative confirmed cases}&
\textbf{Population density}&
\textbf{GDP per-capita (RMB)}\\ \hline
\endhead
\hline \multicolumn{5}{r}{{Continued on next page}} \\ \hline
\endfoot
\hline
\endlastfoot
Chongqing	&Chongqing&	576&	376.43083&	65933\\
Zhejiang&	Wenzhou&	504&	763.832&	65055\\
Guangdong&	Shenzhen&	417	&6484&	189568\\
Beijing&	Beijing&	400	&7794.8&	192957.1\\
Guangdong&	Guangzhou&	346	&2005&	155491\\
Shanghai&	Shanghai&	336	&3814	&126634\\
Henan&	Xinyang&	274&	342.0925&	36951.12\\
Shandong&	Jining&	258&	745.7689&	58972\\
Hunan&	Changsha&	242&	689.9653&	136920\\
Jiangxi	&Nanchang&	229&	749.1894&	95116.22\\
Heilongjiang&	Harbin&	198&	204.4821&	48345.92\\
Anhui&	Hefei&	174	&706.5906&	116352.2\\
Zhejiang&	Hangzhou&	169&	518.958	&140180\\
Anhui&	Bengbu&	160	&640.6788&	50662\\
Henan&	Zhengzhou	&157	&1361.268&	101352.1\\
Zhejiang&	Ningbo&	157	&835.575&	132603\\
Hunan&	Yueyang&	156&	385.9844&	59165\\
Anhui&	Fuyang	&155&	1058.312&	21700\\
Henan&	Nanyang&	155&	376.4511&	35554.64\\
Zhejiang&	Taizhou&	146&	610.846&	79541\\
Sichuan&	Chengdu	&143&	1121.066&	86911\\
Henan&	Zhumadian&	139	&466.5252&	33773.42\\
Tianjin&	Tianjin	&135	&2596.509&	118944\\
Jiangxi&	Xinyu&	130&	373.411&	86791\\
Jiangxi&	Shangrao&	123&	298.7281&	36899\\
Shaanxi	&Xi'an	&120&	989.6814&	86000\\
Jiangxi	&Jiujiang	&118&	260.1498&	55141.93\\
Anhui&	Bozhou&	108&	625.3881	&24388.01\\
Jiangxi&	Yichun&	106&	297.4825&	39268.42\\
Hunan&	Shaoyang	&102&	353.9426&	24178\\
Guangdong&	Zhuhai&	98	&1111.758&	159428\\
Guangdong&	Dongguan&	97&	3404.544&	98939\\
Jiangsu&	Nanjing	&93&	1280.735&	152886\\
Henan&	Shangqiu&	91	&684.3516&	32669.96\\
Jiangsu&	Suzhou&	87	&1238.501&	173765\\
Guangdong&	Foshan&	84&	2043.873&	127691\\
Hong Kong&	Hong Kong&	84&	6733.915&	336147\\
Anhui&	Anqing	&83	&388.8447	&37243.79\\
Hunan&	Changde	&82&	320.3518&	58160\\
Hunan&	Zhuzhou	&80	&357.0236	&65442\\
Jiangsu	&Xuzhou	&79	&748.1513&	76915\\
Jiangxi&	Ganzhou&	76	&219.4027&	40212\\
Henan&	Zhoukou&	76	&725.6292	&30820.63\\
Hunan&	Loudi&	76&	484.3908&	39249\\
Jiangxi&	Fuzhou&	72&	215.0821	&34156.95\\
Fujian&	Fuzhou&	71	&645&	102037\\
Sichuan&	Garz&	69&	7.772442&	22097\\
Anhui&	Lu'an	&69&	338.3381&	24638.49\\
Guangdong&	Zhongshan&	66	&1838.889	&110585\\
Jiangsu&	Huai'an	&66	&491.0269&	73204\\
Guangdong&	Huizhou&	62&	431.25	&85418\\
Shandong&	Qingdao	&60	&674.295&	128459\\
Hunan&	Yiyang	&59	&363.4552&	39937\\
Hebei&	Tangshan&	58	&586.1787&	82692\\
Henan&	Pingdingshan&	58&	797.921	&42586\\
Henan&	Xinxiang&	57	&700.7014&	43696.44\\
Guangxi&	Nanning&	55&	328.0617&	59259\\
Jiangsu&	Wuxi&	55&	1420.899&	174270\\
Zhejiang&	Jinhua&	55&	512.155&	73428\\
Fujian&	Putian&	55	&685.5792&	77325\\
Hainan&	Sanya	&54	&393.8903&	63046.1\\
Yunnan&	Kunming	&53	&326.1905&	76387\\
Henan&	Anyang	&53	&924.4508&	46443.24\\
Heilongjiang&	Shuangyashan&	52	&62.64733&	35527\\
Jiangsu	&Changzhou&	51&	978.6368&	149275\\
Shandong&	Linyi&	49	&617.9618&	44534\\
Jiangsu	&Lianyungang&	48&	593.5653&	61332\\
Hebei	&Cangzhou	&48&	563.0002&	48226\\
Hunan	&Hengyang	&48&	473.1156&	42163\\
Shandong&	Jinan	&47&	912.3639&	106302\\
Shandong&	Yantai	&47&	521.5907&	110231\\
Fujian&	Quanzhou&	47&	790.1907	&97614\\
Heilongjiang&	Suihua&	47	&149.0159&	29625.32\\
Heilongjiang&	Jixi&	46	&74.9566&	27639.35\\
Jilin&	Changchun&	45	&373.3042&	86465\\
Zhejiang&	Jiaxing&	45&	1119.11&	103858\\
Shandong&	Weifang&	44	&580.8031&	65721\\
Guangxi&	Beihai&	44&	525.6818&	73074\\
Hunan&	Yongzhou&	43&	242.9526&	33035\\
Heilongjiang&	Qiqihar&	43&	119.6062&	23676.17\\
Sichuan&	Dazhou&	42&	342.6378&	28066\\
Zhejiang&	Shaoxing&	42&	608.1652&	107853\\
Anhui&	Suzhou&	41&	580.5048&	28693.98\\
Jiangsu&	Nantong	&40	&692.9567&	115320\\
Hunan&	Huaihua&	40&	180.6035&	30449\\
Sichuan	&Nanchong&	39	&512.8576	&28516\\
Henan&	Xuchang&	39&	888.1906&	63987.61\\
Hainan&	Haikou	&39&	973.3623&	56055.67\\
Hunan&	Chenzhou	&39	&244.7258&	50482\\
Anhui&	Ma'anshan	&38&	577.1796&	82075\\
Shandong&	Weihai&	38	&488.1835&	128774\\
Shandong&	Liaocheng&	38	&696.5371	&51935\\
Shandong&	Dezhou&	37	&560.7567&	58252\\
Shanxi&	Jinzhong&	37	&206.2	&42916\\
Jiangsu	&Taizhou&	37	&801.0181&	110180\\
Hunan&	Xiangtan&	36&	572.2733&	75609\\
Gansu&	Lanzhou	&36	&284.3636&	73042\\
Guizhou&	Guiyang	&36	&606.9471&	78449\\
Shandong&	Tai'an&	35	&726.7105&	64714\\
Henan&	Luohe&	35&	1020.252&	46318\\ \label{table3}
\end{longtable}

\clearpage
\newpage

\begin{longtable}{lllll}
\caption{The pool of cities in the group of Wenzhou and their corresponding weights}\\

\hline \multicolumn{1}{l}{\textbf{Province}} &
\multicolumn{1}{l}{\textbf{City}} &
\multicolumn{1}{l}{\textbf{Population density}}& \multicolumn{1}{l}{\textbf{GDP per-capita (RMB)}}&
\multicolumn{1}{l}{\textbf{Weight}}\\ \hline
\endfirsthead
\multicolumn{5}{c}%
{{\bfseries \tablename\ \thetable{} -- continued from previous page}} \\
\hline \multicolumn{1}{l}{\textbf{Province}} &
\multicolumn{1}{l}{\textbf{City}} &
\multicolumn{1}{l}{\textbf{Population density}}& \multicolumn{1}{l}{\textbf{GDP per-capita (RMB)}}&
\multicolumn{1}{l}{\textbf{Weight}}\\ \hline
\endhead
\hline \multicolumn{5}{r}{{Continued on next page}} \\ \hline
\endfoot
\hline
\endlastfoot
Chongqing	&Chongqing&	376.43083&	65933 &0\\
Shandong&	Jining&	745.7689&	58972&	0\\
Jiangxi&	Nanchang&	749.1894&	95116.22&	0\\
Anhui&	Hefei&	706.5906&	116352.2&	0\\
Henan&	Zhengzhou&	1361.268&	101352.1&	0\\
Hunan&	Yueyang&	385.9844&	59165&	0\\
Sichuan&	Chengdu&	1121.066&	86911&	0\\
Tianjin&	Tianjin&	2596.509&	118944&	0\\
Jiangxi&	Xinyu&	373.411&	86791&	0.69\\
Shaanxi&	Xi'an&	989.6814&	86000&	0\\
Jiangxi&	Jiujiang&	260.1498&	55141.93&	0\\
Guangdong&	Dongguan&	3404.544&	98939&	0\\
Hunan&	Changde&	320.3518&	58160&	0\\
Hunan&	Zhuzhou&	357.0236&	65442&	0\\
Jiangsu&	Xuzhou&	748.1513&	76915&	0\\
Fujian&	Fuzhou&	645&	102037&	0\\
Guangdong&	Zhongshan&	1838.889&	110585&	0\\
Jiangsu&	Huai'an&	491.0269&	73204&	0\\
Guangdong&	Huizhou&	431.25&	85418&	0\\
Hebei&	Tangshan&	586.1787&	82692&	0\\
Guangxi&	Nanning&	328.0617&	59259&	0\\
Zhejiang&	Jinhua&	512.155&	73428&	0\\
Fujian&	Putian&	685.5792&	77325&	0\\
Hainan&	Sanya&	393.8903&	63046.1&	0.31\\
Yunnan&	Kunming&	326.1905&	76387&	0\\
Jiangsu&	Lianyungang&	593.5653&	61332&	0\\
Shandong&	Jinan&	912.3639&	106302&	0\\
Shandong&	Yantai&	521.5907&	110231&	0\\
Fujian&	Quanzhou&	790.1907&	97614&	0\\
Jilin&	Changchun&	373.3042&	86465&	0\\
Zhejiang&	Jiaxing&	1119.11&	103858&	0\\
Shandong&	Weifang&	580.8031&	65721&	0\\
Guangxi&	Beihai&	525.6818&	73074&	0\\
Zhejiang&	Shaoxing&	608.1652&	107853&	0\\
Jiangsu&	Nantong&	692.9567&	115320&	0\\
Henan&	Xuchang&	888.1906&	63987.61&	0\\
Hainan&	Haikou&	973.3623&	56055.67&	0\\
Anhui&	Ma'anshan&	577.1796&	82075&	0\\
Shandong&	Dezhou&	560.7567&	58252&	0\\
Jiangsu&	Taizhou&	801.0181&	110180&	0\\
Hunan&	Xiangtan&	572.2733&	75609&	0\\
Gansu&	Lanzhou&	284.3636&	73042&	0\\
Guizhou&	Guiyang&	606.9471&	78449&	0\\
Shandong&	Tai'an&	726.7105&	64714&	0\\ \label{table4}
\end{longtable}

\clearpage
\newpage
\setcounter{figure}{0}
\begin{figure}[H]
\begin{center}
\includegraphics[totalheight=5.4in, width=6.8in, origin=c]{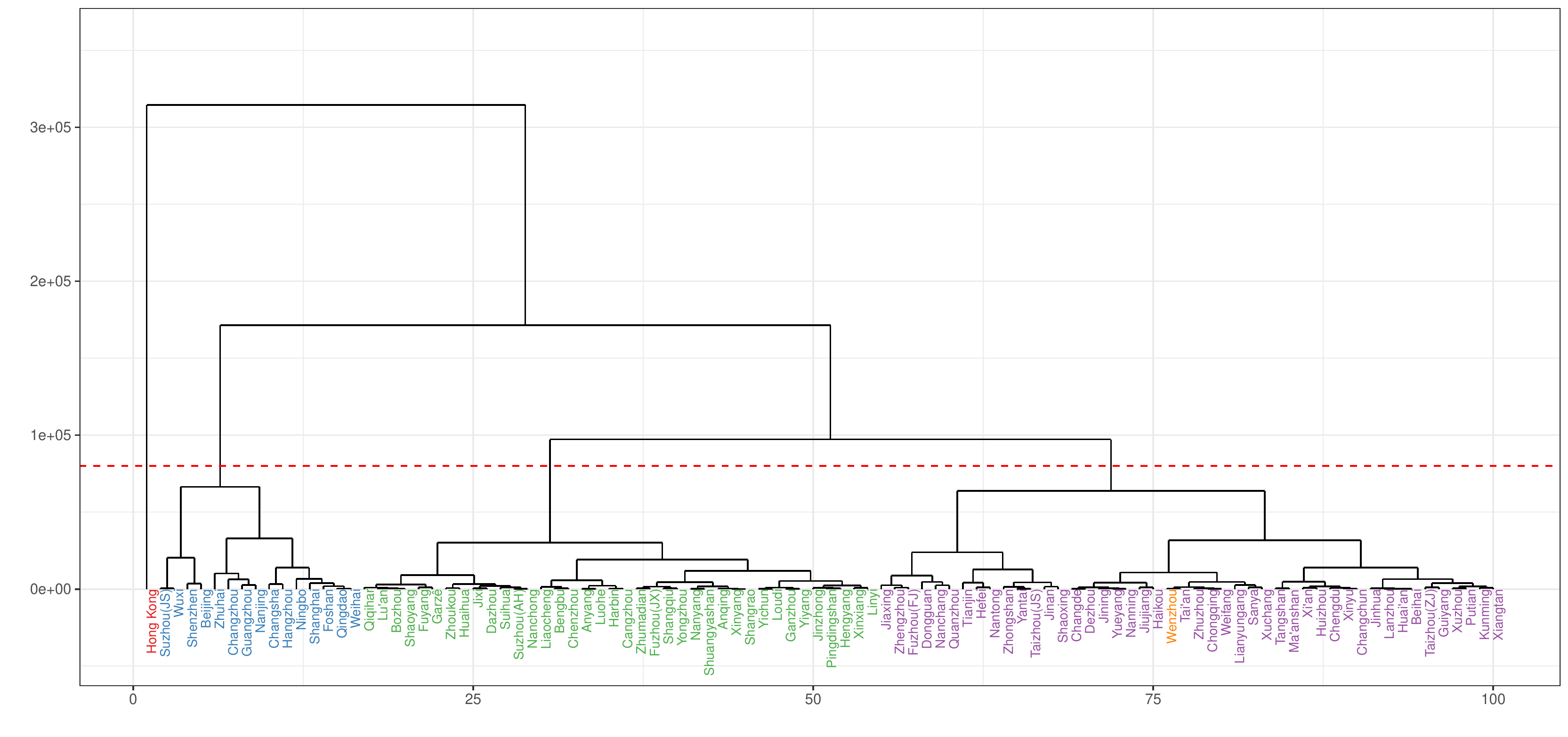}
\end{center}
\caption{\textbf{A dendrogram tree of clustering results}} \label{fig5}
\end{figure}

\clearpage
\newpage
\begin{figure}[H]
\begin{center}
\includegraphics[totalheight=5.4in, width=6.8in, origin=c]{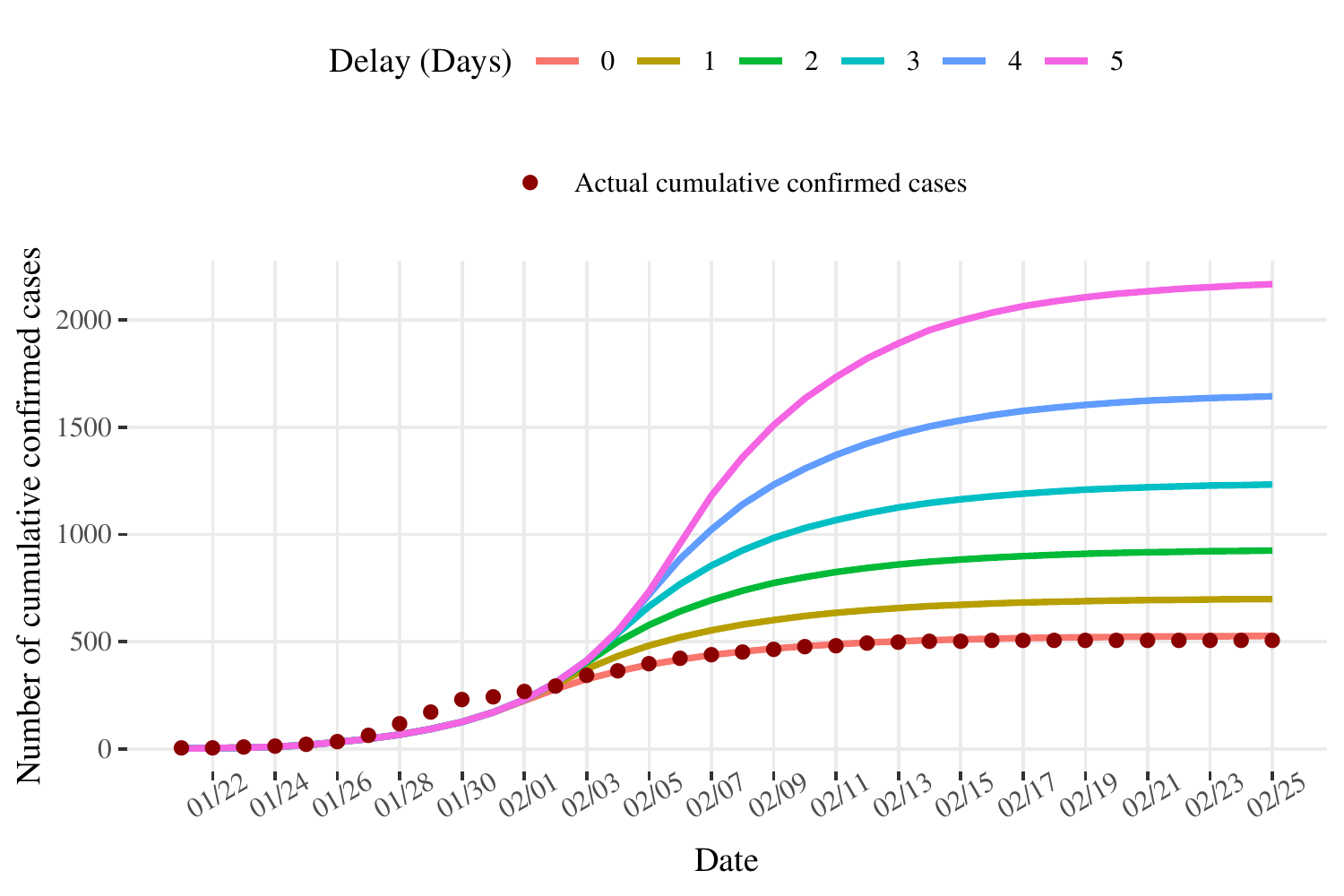}
\end{center}
\caption{\textbf{The expected number of cumulative confirmed cases for stringent interventions if they were delayed by different days} The points represent the actual cumulative confirmed cases of Wenzhou.} \label{fig6}
\end{figure}

\clearpage
\newpage
%\section*{Acknowledgments}
%%The authors thank the editor, the associate editor and two anonymous reviewers for valuable comments that have remarkably improved the manuscript.
%
%\section*{Funding}
%Tian's research was supported by Fundamental Research Funds for the Central Universities (Grant No. 19lgpy236). Wen's research is partially supported by NSFC (Grant No. 11801540), Natural Science Foundation of Guangdong (Grant No. 2017A030310572) and Fundamental Research Funds for the Central Universities (Grant No. WK2040170015, WK2040000016). Wang's research was partly supported by NSFC(Grant No. 11771462, Grant No. 71991474), National Key Research and Development Program of China (Grant No. 2018YFC1315400) and Key Research and Development Program of Guangdong, China (Grant No. 2019B020228001).

\newpage
\bibliographystyle{ECA_jasa}
\bibliography{meantest}

\end{document}